\documentclass[aps,prb,reprint,showpacs,floatfix]{revtex4-1}  
\usepackage{amsmath,amssymb,xcolor}
\usepackage[colorlinks,bookmarks=false,citecolor=blue,linkcolor=blue,urlcolor=blue]{hyperref}
\usepackage{graphicx,epstopdf}
\usepackage{array,graphicx,multirow}
\newcolumntype{C}[1]{>{\centering\arraybackslash}m{#1}}
\begin{document}
\title{Topological approach to quantum liquid ground states on geometrically frustrated Heisenberg antiferromagnets}
\author{Santanu Pal}
\email{20001575@iitb.ac.in}
\altaffiliation[Present address: ]
{Department of Physics, Indian Institute of Technology Bombay, Mumbai, MH 400076, India}
\affiliation{Department of Physical Sciences, Indian Institute of Science Education and Research-Kolkata, W.B. 741246, India}

\author{Anirban Mukherjee}
\email{am14rs016@iiserkol.ac.in}
\author{Siddhartha Lal}
\email{slal@iiserkol.ac.in}
\affiliation{Department of Physical Sciences, Indian Institute of Science Education and Research-Kolkata, W.B. 741246, India}
\begin{abstract}
We have formulated a twist operator argument for the geometrically frustrated quantum spin systems on the kagome and triangular lattices, thereby extending the application of the Lieb-Schultz-Mattis (LSM) and Oshikawa-Yamanaka-Affleck (OYA) theorems to these systems.
The equivalent large gauge transformation for the geometrically frustrated lattice differs from that for non-frustrated systems due to the existence of multiple sublattices in the unit cell and non-orthogonal basis vectors.
Our study for the $S=1/2$ kagome Heisenberg antiferromagnet at zero external magnetic field gives a criterion for the existence of a two-fold  degenerate ground state with a finite excitation gap and fractionalized excitations. At finite field, we predict various plateaux at fractional magnetisation, in analogy with integer and fractional quantum Hall states of the primary sequence. These plateaux correspond to gapped quantum liquid ground states with a fixed number of singlets and spinons in the unit cell.
A similar analysis for the triangular lattice predicts a single fractional magnetization plateau at $1/3$. Our results are in broad agreement with numerical and experimental studies.
\end{abstract}
\pacs{75.,75.10.Jm,75.10.Kt,75.50.Ee,75.78.-n}
\maketitle
\section{Introduction}
\noindent
Frustrated spin systems have, for several decades, drawn significant attention in the search for exotic ground states. The causes of frustration are several~\cite{anderson1973resonating,anderson1987resonating,PhysRevB.45.7832,kitaev2006anyons}, with special emphasis given to lattices on which the classical N\'eel ground states of the nearest neighbour (n.n) Heisenberg antiferromagnet cannot be stabilised due to an intrinsic frustration. The kagome and triangular lattices in 2D and the pyrochlore lattice in 3D are classic examples of such systems. A large number of theoretical as well as experimental studies have sought novel ground states such as spin liquids and spin ice~\cite{lee2008end,balents2010spin,RevModPhys.88.041002}, as well as states possessing topological order and fractionalized excitations~\cite{han2012fractionalized}. In spite of extensive studies on the $S=1/2$ Heisenberg kagome antiferromagnet (HKA), the nature of the ground state and the existence of a spectral gap remain inconclusive. Some studies support the existence of a gap and short-ranged resonating valence bond (RVB) order~\cite{PhysRevX.4.011025,PhysRevB.95.235107,jiang2012identifying,fu2015evidence,yan2011spin,PhysRevLett.97.207204}, while others suggest a gapless spectrum and algebraic order~\cite{PhysRevLett.98.117205,PhysRevB.75.184406,PhysRevB.84.020407,PhysRevB.87.060405,PhysRevB.89.020407,PhysRevLett.118.137202,PhysRevX.7.031020}. Another interesting aspect of geometrically frustrated spin systems is that they can possess nontrivial plateaux at zero and fractional magnetisation~(see, e.g., \cite{schulenburg2002macroscopic,honecker2004magnetization,schnack2018magnetism,zhitomirsky2004exact,nakano2010magnetization,chubokov1991quantum,hida2001magnetization,nishimoto2013controlling,PhysRevLett.114.227202,PhysRevLett.102.137201,PhysRevB.67.104431} for triangular and kagome lattices). The existence of such plateaux indicates a finite gap in the energy spectrum and the possibility of ground states with non-trivial topological features analogous to the quantum Hall effects~\cite{PhysRevLett.78.1984,PhysRevB.94.134410,PhysRevB.92.094433,PhysRevB.30.1097}. In fact, the ground state wavefunction for the plateau at fractional magnetization $m=7/9$ is known exactly~\cite{schulenburg2002macroscopic,zhitomirsky2004exact,changlani2019resonating}.

\par
There exist very few methods that, relying solely on the symmetries of the Hamiltonian, can offer qualitative insight on the nature of the ground state and the low-energy excitation spectrum. One of these is the Lieb-Schultz-Mattis (LSM) theorem\cite{lieb1961two}.  Originally formulated for the spin-$1/2$ n.n. Heisenberg antiferromagnet chain, it was extended to higher dimensions for geometrically non-frustrated systems more recently~\cite{PhysRevB.37.5186,PhysRevLett.84.1535,PhysRevB.69.104431}.
The theorem relates the existence (or lack) of a spectral gap to the sensitivity of the ground state to adiabatic changes in boundary conditions implemented by a twist operator. 
A degeneracy of the ground state can also be gauged from the non-commutativity between the lattice translation and twist operators. 
Recent works have been devoted to extending the applicability of the LSM theorem to systems with a variety of interactions (e.g., extended, anisotropic, bond-alternating, Dzyaloshinskii-Moriya and even frustrating)~\cite{Nomura,isoyama2017discrete,tasaki2018lieb}. This is in broad agreement with some numerical studies of (quasi-)one dimensional systems (e.g., chains and ladders)~\cite{PhysRevB.94.224421,antkowiak2017universal,florek2018lieb}. 
These works indicate that the minimum requirements for the LSM theorem are spin Hamiltonians possessing $U(1)$ spin symmetry, translation invariance in real space and short-ranged interactions. Importantly, without assuming either a bipartite lattice or a unique ground state, Ref.(\cite{Nomura}) extends the LSM theorem to frustrated spin systems in quasi-one dimension where ground states may be degenerate. Further, Oshikawa {\it et al.}
~\cite{PhysRevLett.78.1984} extended the LSM-theorem to the case of finite magnetization (the OYA criterion), using which one can predict possible magnetization plateaux for finite external magnetic field. 
It is important to note that the OYA-criterion has been extended to quantum antiferromagnetic systems in abitrary spatial dimensions by Tanaka {\em et al.}~\cite{tanaka2009geometric,tanaka2015short} with the help of effective field theory and renormalisation group (RG) analyses.
Further, the OYA-criterion has been successful in predicting plateaux for the $S=1/2$ HKA~\cite{fledderjohann1999soft,PhysRevB.88.144416,nishimoto2013controlling}. Very recently, two of us have predicted possible magnetization plateau states in $S=1/2$ pyrochlore lattice by using a similar formalism to that presented here\cite{PhysRevB.100.104421}. In a RG analysis of the $S=1/2$ HKA on the kagome lattice~\citep{pal2019}, we have also shown that the twist operator we present here is responsible for the formation of the spectral gap that protects the $1/3$ magnetization plateau ground state. This provides important evidence for the origin of the spectral gap assume in the twist operator based analysis.
\par  
As presented in Section \ref{twistopsection}, the main goal of the present work is to define the twist operator (also called a large gauge transformation operator~\cite{PhysRevLett.84.1535,PhysRevB.70.245118}) for geometrically frustrated 2D  lattices (e.g., kagome and triangular). The subtlety in the form of the twist operator in such lattices lies in identifying the non-trivial unit cell and the associated basis vectors. Then, from the usual non-commutativity between twist and translation operators, we obtain the possibility of gapped, doubly-degenerate ground states with interpolating fractional excitations
for the HKA at zero field in Section \ref{lsmoyasec}. 
Further, in sections \ref{lsmoyasec} and \ref{trisec}, we demonstrate the existence of several plateaux at finite magnetisation from an OYA-like criterion on the kagome and triangular lattices. These compare favourably with results obtained from various numerical methods~\cite{nishimoto2013controlling}.
The non-saturation plateaux obtained at non-zero field from such spectral flow arguments correspond to quantum liquid ground states in which the unit cells comprise of short-ranged RVBs along with a fixed number of spinon excitations~ \cite{PhysRevLett.81.4484,PhysRevB.92.094433}. This should be contrasted with proposals of quantum solid valence bond solid (VBS) ground states \cite{PhysRevLett.59.799} and $SU(2)$ symmetry broken classical ground states~\cite{PhysRevB.45.12377} for geometrically frustrated 2D spin systems. We conclude in Section \ref{consec}, presenting some open directions. For the sake of completeness, we present the details of the calculations for the energy cost related to the twist operation and the LSM-like theorem for the kagome lattice in Appendices \ref{EnergyCost} and \ref{LSMcalculation}. 
\begin{figure}[!ht]
\includegraphics[scale=.2]{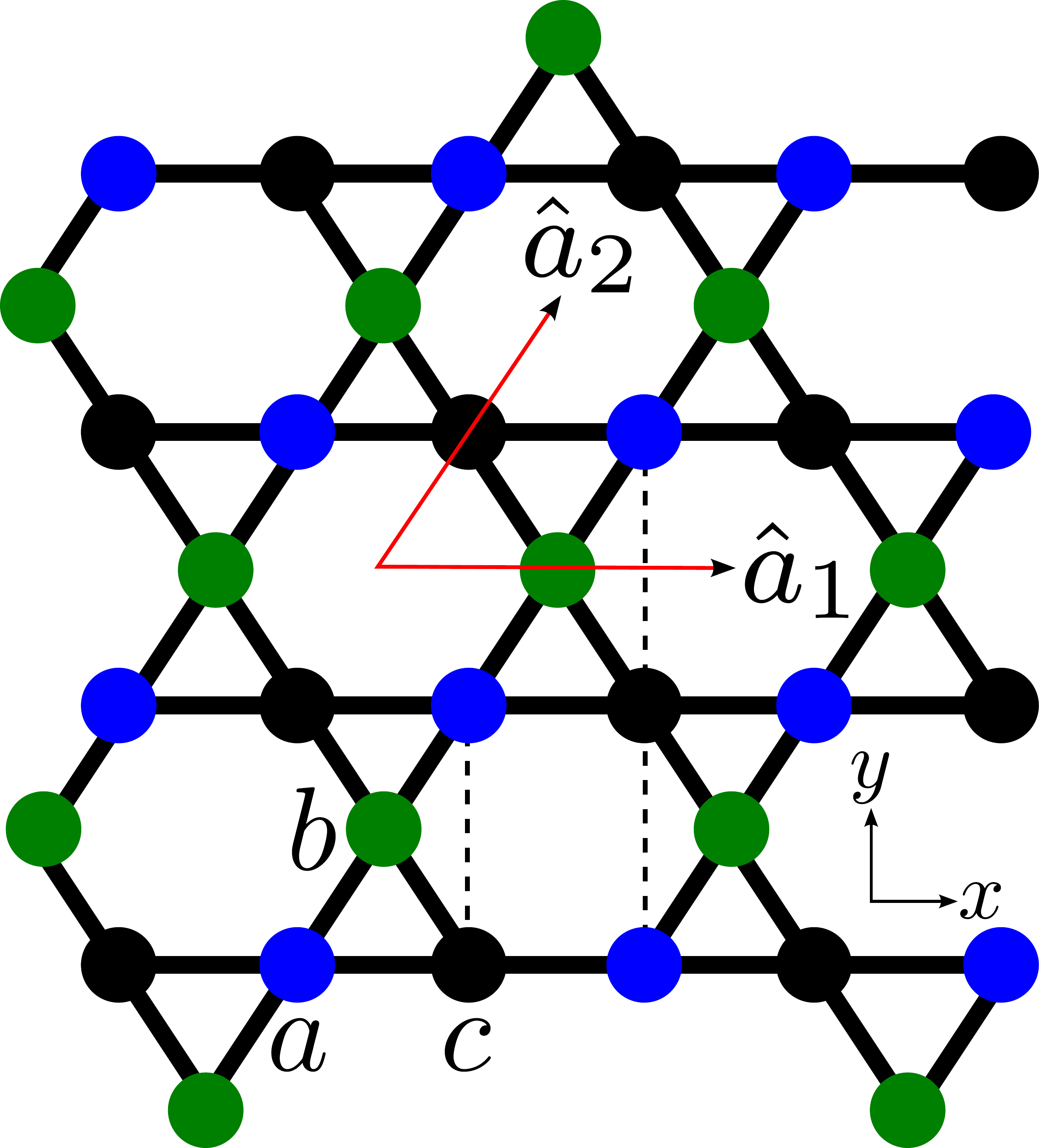}
\caption{Schamatic diagram of kagome lattice with the  basis vectors $\hat{a}_{1}=\hat{x}, \hat{a}_{2}=\frac{1}{2}\hat{x}+\frac{\sqrt{3}}{2}\hat{y}$, so the distance between nearest neighbour sites is half. Every triangular unit cell has three different sublattice labelled by $a, b$ and $c$ ( blue, green and black respectively). The dashed lines show the non-zero projection of sites in the $\hat{a}_2$ direction along $\hat{a}_1$. }
\label{kagomeunitcell}
\end{figure}
\section{Twist Operator for the kagome lattice}\label{twistopsection}
\noindent
The kagome system has two basis vectors $\hat{a}_1$ and $\hat{a}_{2}$ with which the complete lattice can be spanned [Fig.(\ref{kagomeunitcell})]. The Hamiltonian for $S=\frac{1}{2}$ n.n HKA in a field $h$ is~\cite{PhysRevB.77.224413} 
\begin{eqnarray}
H=J \sum_{<\vec{r}\vec{r}'>} \vec{S}_{\vec{r}}\cdot \vec{S}_{\vec{r}'} - h\sum_{\vec{r}} \hat{S}^{z}_{\vec{r}}~,
\label{Hamiltonian}
\end{eqnarray}
where the spin exchange $J>0$ and sum is over n.n sites. Here $\vec{r}\in (\vec{R},j)$, with $\vec{R}=n_1\hat{a}_1+n_2\hat{a}_2$ ($n_1, n_2$ are integer) the lattice vector for a three sub-lattice  unit cell (up triangles) and $j\in{(a, b, c)}$ are the three sub-lattices. 
For $N_1$ and $N_2$ being the number of each sub-lattice along the $\hat{a}_1$ and $\hat{a}_2$ directions respectively, the total number of sites in the lattice is $3N_1 N_2$. Below, we will consider periodic boundary conditions (PBC) along $\hat{a}_1$ direction. Now, for $\delta$ being the distance between n.n sites, $L_{\hat{a}_1}=2\delta N_1$ and $L_{\hat{a}_2}=2\delta N_2$ are the lengths along the $\hat{a}_1$ and $\hat{a}_2$ directions respectively. Hereafter, we will consider $\delta=1$.
\par
In the LSM theorem~\cite{lieb1961two}, a {\it twist} (i.e., a change in boundary conditions) is equivalent to insertion of an Aharanov-Bohm (AB) flux~\cite{PhysRevLett.84.1535,PhysRevB.70.245118} that generates a vector potential along the periodic direction. This is analogous to Laughlin's flux insertion for the quantum Hall effect~\cite{PhysRevB.23.5632}. By this argument, one can extend the LSM theorem to higher dimensions~\cite{PhysRevB.69.104431}, with twisting equivalent to a large gauge transformation of the Hamiltonian. We expect an invariance of the spectrum under a large gauge transformation equivalent to the adiabatic insertion of a full flux quanta ($2\pi$, in units $h=c=e=1$). The twisted wavefunction, however, reveals the effect of the flux. Thus, we compute a shift in the crystal momentum 
by applying a gauge transformation that reverses precisely the shift in the eigenspectrum due to the flux~\cite{PhysRevB.70.245118}. This shift is revealed by a non-commutativity between the translation and twist operators. 
\par
In applying the LSM theorem on geometrically frustrated lattices, one has to be careful in defining a suitable large gauge transformation. 
On such lattices, the basis vectors are usually not orthogonal to one another (see Fig.(\ref{kagomeunitcell}) for the kagome lattice). Therefore, spins at different sites along a basis vector (other than that along which the twist is applied) differ in the phase induced by the equivalent AB flux. We place the system shown in Fig.(\ref{kagomeunitcell}) on a cylinder, with PBC along $\hat{x}\equiv\hat{a}_1$. 
Now if we apply an AB-flux along the axis of the cylinder, a time-varying vector potential will be induced along $\hat{a}_1$ direction. 
For a uniform gauge $A(x)=2\pi/L_{\hat{a}_1}$ and $A(y)=0$, there will be no change in the phase of spins on sites with the same $y$-coordinate. Given that $\hat{a}_{2}$ does not coincide with $\hat{y}$, 
the phase acquired by the spins varies along $\hat{a}_{2}$. 
Below, we account for this subtlety in constructing twist operators for the 
kagome and triangular lattices. 
\par 
Given that $[S^\alpha_{\vec{r}},S^\beta_{\vec{r}'}]=0$ for $\vec{r}\neq\vec{r}'$, where $\alpha, \beta \in \{x,y,z\}$, we can define separate twist operators for the three sub-lattices ($\hat{O}_a$, $\hat{O}_b$ and $\hat{O}_c$) and combine them for the complete twist operator $\hat{O}=\hat{O}_a\hat{O}_b\hat{O}_c$~.
Then, for a flux quantum along $\hat{y}$, the phase difference between spins belonging to the nearest sites of the same sub-lattice and with fixed $n_2$ ($n_1$) coordinate is given by $2\pi/N_1$ ($\pi/N_1$); see dashed lines in Fig.(\ref{kagomeunitcell}). 
Therefore, with the site marked as $a$ in Fig.(\ref{kagomeunitcell}) chosen as the reference site, 
the twist operator for sub-lattice $a$ ($\hat{O}_{a}$) is given by
\begin{eqnarray}
\hat{O}_a=\exp \big[i\frac{2\pi}{N_1}\sum_{\vec{R}}(n_1+\frac{n_2}{2}) \hat{S}^z_{\vec{R},a}\big]~.
\end{eqnarray}
In a given unit cell, the phases acquired by $b$ and $c$ sub-lattices differ by $\frac{1}{4}(2\pi/N_1) $ and $\frac{1}{2}(2\pi/N_1) $ respectively with respect to the $a$ sub-lattice. Thus, the twist operator for sub-lattice $b$ is given by
\begin{eqnarray}
\hat{O}_b &=& \exp \big[i\frac{2\pi}{N_1}\sum_{\vec{R}}(n_1+\frac{n_2}{2}+\frac{1}{4}) \hat{S}^z_{\vec{R},b}\big]~,
\end{eqnarray}
while $\hat{O}_{c}$ is identical in form, with only the term proportional to $1/4$ in the exponent replaced by one proportional to $1/2$. Combining the three, we obtain the complete twist operator for kagome lattice
\begin{eqnarray}
\hspace*{-0.2cm}\hat{O}=\hspace*{-0.05cm}\exp\hspace*{-0.05cm} \big[i\frac{2\pi}{N_1}\Big(\hspace*{-0.1cm}\sum_{\vec{r}}\hspace*{-0.1cm}(n_1\hspace*{-0.1cm}+\hspace*{-0.1cm}\frac{n_2}{2}) \hat{S}^z_{\vec{r}}+\hspace*{-0.1cm}\sum_{\vec{R}}(\frac{1}{4} \hat{S}^z_{\vec{R},b}\hspace*{-0.1cm}+\hspace*{-0.1cm}\frac{1}{2} \hat{S}^z_{\vec{R},c})\Big)\big].
\label{twistoperator}
\end{eqnarray}
This form of the twist operator differs from that obtained for non-frustrated lattices~\cite{PhysRevLett.84.1535,PhysRevB.70.245118} in two ways. The term proportional to $n_{2}$ appears due to the non-orthogonality of the basis vectors, while the terms proportional to $\hat{S}^z_{\vec{R},b}$ and $\hat{S}^z_{\vec{R},c}$ arise due to the different phase twists acquired by the sub-lattices of the kagome system. 
We will use this twist operator to obtain the nature of the ground state and low-energy spectrum for the HKA. In Appendix \ref{EnergyCost}, 
we show that the excitation gap between the ground state and the twisted state vanishes in the thermodynamic limit for a vanishing spin stiffness~\cite{PhysRevB.69.104431,misguich2002degeneracy}.
\section{LSM-like theorem and OYA-like criterion for the kagome lattice}\label{lsmoyasec}
\noindent
We denote the unit translation operator along $\hat{a}_1$ direction as $\hat{T}_{\hat{a}_1}$,
such that $\hat{T}_{\hat{a}_1} \hat{S}^z_{n_1,n_2,j}\hat{T}^\dagger_{\hat{a}_1}
=\hat{S}^z_{n_1+1,n_2,j}$~. For PBC along $\hat{a}_1$ direction, 
we obtain the identity (see Appendix \ref{LSMcalculation} 
for a detailed calculation)
\begin{eqnarray}
\hat{T}_{\hat{a}_1}\hat{O}\hat{T}^\dagger_{\hat{a}_1}= \hat{O}\exp\big[-i\frac{2\pi}{N_1}(\hat{S}^z_{Tot}-N_1 N_2 \hat{S}^z_{\triangle})\big]~,
\label{LSMcriterian}
\end{eqnarray}
$N_{2}\hat{S}^z_\triangle$ is the $z$-component of the vector sum of all spins within the $N_{2}$ unit cells 
where the total magnetization is given by $\hat{S}^z_{Tot}=\sum_{\vec{r}}\hat{S}^z_{\vec{r}}$. We obtain the factor $N_{2}\hat{S}^z_\triangle$ as the $z$-component of the vector sum of all spins within the $N_{2}$ unit cells lying on a line along $\hat{a}_{2}$ (the {\it boundary} line~\cite{PhysRevB.70.245118}) by assuming translation invariance along that direction. For the kagome lattice, 
$S_{\triangle}=1/2,3/2$ such that the eigenvalues of $\hat{S}^{z}_{\triangle}$ are $\pm 1/2,\pm 3/2$. As mentioned earlier, the applicability of the LSM theorem demands a $U(1)$ invariance of the ground state, i.e., it is labelled by the eigenvalue of $\hat{S}^{z}_{Tot}$. For the case of $h=0$, the total number of sites in the lattice ($N_{1}\times N_{2}$) has to be even in order to guarantee the time reversal invariance of the ground state, i.e., $\hat{S}^{z}_{Tot}|\psi_{0}\rangle =0$.
\par  
Then, at zero field, the matrix element arising from eqn.(\ref{LSMcriterian}) becomes 
\begin{eqnarray}
\langle \psi_0|\hat{T}_{\hat{a}_1}\hat{O}\hat{T}^\dagger_{\hat{a}_1}|\psi_0\rangle= \langle \psi_0|\hat{O}\exp\big[-i N_2~(\text{mod} ~\pi)\big]|\psi_0\rangle.
\label{LSMsinglet}
\end{eqnarray}
For $N_2\in$ odd and the lowest excited state $|\psi_1\rangle=\hat{O}|\psi_0\rangle$, eqn.(\ref{LSMsinglet}) leads to $\langle \psi_0|\psi_1\rangle=0$,
i.e., the ground state and the lowest lying excited state are orthogonal to one another. Therefore, employing the LSM argument used for the $S=1/2$ Heisenberg chain as well as ladder systems ~\cite{lieb1961two,PhysRevLett.81.4484,PhysRevB.65.153110}, we find that the $S=1/2$ HKA can have one of two possible ground states. The first possibility is that, without the breaking of any symmetries, there exists a many-body gap separating the excitation spectrum from a two-fold degenerate ground state. This is in agreement with the finding of a small zero-magnetization plateau from numerical investigations of the HKA in Ref.(\cite{nishimoto2013controlling}). These two ground states are topologically separated from one another: the AB flux threading is equivalent to the insertion of a \emph{vison} carrying a crystal momentum $\pi$ into the hole of the cylinder~\cite{PhysRevB.70.245118}. 
This is the signature of a $Z_2$ fractionalised insulating phase~\cite{PhysRevB.70.245118,PhysRevB.65.024504,PhysRevB.62.7850}. The degeneracy in the ground state manifold appears in the thermodynamic limit, along with a  spin stiffness that decays exponentially with system size~\cite{misguich2002degeneracy,PhysRevLett.81.4484}. This justifies the adiabatic insertion of the AB flux over timescales much longer than the inverse gap~\cite{PhysRevLett.84.1535,PhysRevB.69.104431,PhysRevB.70.245118}.
The other possibility is that, in the thermodynamic limit, the excitation spectrum generated by $\hat{O}$
collapses, causing the many body gap to vanish. Indeed, another recent work suggests a $U(1)$ gapless spin liquid ground state in the HKA~\cite{chen2018thermodynamics}. Thus, the LSM-like arguments presented above are, by construction, unable to resolve between these two possibilities. On the other hand, for $N_2\in$ even, $\langle \psi_0|\psi_1\rangle\neq 0$ and the approach taken here does not yield any firm conclusions about the presence of a gap or ground state degeneracy. 
\par
We will now focus on the properties at non-zero magnetic field.
Defining magnetization per site as $m=S^z_{Tot}/3N_1N_2$, eqn.(\ref{LSMcriterian}) becomes 
\begin{eqnarray}
\hat{T}_{\hat{a}_1}\hat{O}\hat{T}^\dagger_{\hat{a}_1}= \hat{O}\exp\big[-i2\pi (3N_2)(m-\frac{\hat{S}^z_\triangle}{3})\big]~.
\label{OYA}
\end{eqnarray}
The appearance of magnetisation plateaux can be understood by noting that we can write the odd integer $N_{2}$ as the product of two odd numbers, $N_{2}=(2p+1)(2q+1)$ where $(p,q)$ can be zero or any positive integer. Then, denote $3N_{2}= Q_{m}(2q+1)$, where $Q_{m}=3(2p+1)$ corresponds to the size of a {\it magnetic} unit cell. The fundamental unit cell of the kagome lattice (see Fig.(\ref{kagomeunitcell})) has $p=0$ and $Q_{m}=3$ spins, whereas the simplest {\it enlarged} unit cell has $p=1$ and $Q_{m}=9$ spins.
We can then derive the OYA-like criterion from eqn.(\ref{OYA}) in terms of the fractional magnetisation, $m/m_{s}$ (where $m_{s}=1/2$ is the saturation magnetisation per site), by requiring that the argument of the exponential is an integer $n$ (upto a factor of $2\pi (2q+1)$).   This is in analogy with the integer quantum Hall effect~\cite{PhysRevLett.78.1984}.
Thus, we obtain
\begin{eqnarray}
\frac{Q_m}{2}(\frac{m}{m_s}-\frac{1}{3})=n \quad\text{or}\quad \frac{Q_m}{2}(\frac{m}{m_s}-1)=n~,
\label{OYA1}
\end{eqnarray}
for $S^{z}_{\triangle}=1/2$ 
and 
for $S^{z}_{\triangle}=3/2$ respectively. 
\begin{table}[h]
\begin{center}
\renewcommand{\arraystretch}{1.5}
\begin{tabular}{|C{1cm}|C{2cm}|C{2cm}|C{2cm}|}\hline
\textbf{$Q_m$} & \textbf{$m/m_s$} &  \textbf{$S^z_{\triangle}=1/2$} &  \textbf{$S^z_{\triangle}=3/2$}\\ \hline
\ 3 &  $1/3$ &  $n=0$ &  $n=-1$ \\ \hline
\multirow{4}{*}{9} &  $1/9$ &  $n=-1$ &  $ n=-4$ \\ \cline{2-4}
  & $1/3$ & $n=0$ & $ n=-3$ \\ \cline{2-4}
  & $5/9$ & $n=1$ & $ n=-2$ \\ \cline{2-4}
  & $7/9$ & $n=2$ & $ n=-1$ \\ \hline
\end{tabular}
\end{center}
\caption{Plateaux in the fractional magnetization ($m/m_{s}$) and the corresponding $(n, S^{z}_{\triangle}, Q_{m})$ values in eqn.(\ref{OYA1}). The symbols are defined in the text.}
\label{Table}
\end{table}
\par\noindent
The table (\ref{Table}) indicates the positions of various plateaux at fractional magnetisation in the HKA. The location of the plateaux agree with results obtained from numerical and experimental works~\cite{nishimoto2013controlling,PhysRevB.88.144416,PhysRevLett.114.227202}. Motivated by Ref.(\cite{PhysRevLett.78.1984}), equn.(\ref{OYA1}) reveals an analogy between the magnetisation plateaux for $S^z_{\triangle}=3/2$ and quantum Hall ground states. For instance, the plateau $m/m_{s}=1/3$ state arising from a fundamental unit cell ($Q_m=3$) is analogous to the integer quantum Hall (IQH) state with filling factor $\nu=1$. This argument extends to a unit cell enlargement of $Q_{m}=3(2p+1)$, e.g., the four plateaux arising from the three-fold enlargement ($Q_m=9$) are in analogy with fractional quantum Hall (FQH) states with $\nu=|n|/Q_m$~\cite{PhysRevB.30.1097}. 
Further, 
these ground states 
contain a fixed number of spinon excitations and RVB singlets~\cite{PhysRevLett.84.3370}: 
the fractional magnetisation $m/m_{s}$, the quantity ($Q_m m/m_s)$ and $|n|$ correspond to the spinon density, spinon number and number of singlets within the magnetic unit cell respectively. 
\par
We now turn to the plateaux obtained for $S_{\triangle}=1/2=S^z_{\triangle}$. The wavefunctions of an isolated triangle of three spin-1/2s (a fundamental unit cell) in the $S_{\triangle}=1/2=S^z_{\triangle}$ sector involve linear combinations of states composed of direct products of a given spin-1/2 and the singlet and triplet states of the other two spin-1/2s  
(see, e.g., eqn. (16) of Ref.(\cite{dai2004classical}). Then, the $1/3$ plateau in $Q_m=3$ can be seen to arise from wavefunctions composed entirely of linear combinations of direct products of single spin-1/2 and triplet states of the other two spins (i.e., a singlet bond count for the fundamental unit cell being $|n|=0$).
For the three-fold enlarged unit cell of $Q_m=9$,  
the $1/9$ and $5/9$ plateau states possess a wavefunction in which one of the three triangles involves a singlet ($|n|=1$). Similarly, the $1/3$ state has a wavefunction with no singlets in any of the three triangles, while the $7/9$ state has singlets in any two triangles. 
\section{Magnetisation plateaux for the triangular lattice}\label{trisec}
\noindent 
We now extend our analysis to the triangular lattice.
Although the triangular lattice possesses geometrical frustration, it has a simple unit cell with an invariance of the Hamiltonian due to translation by one lattice site. Further, it has two basis vectors identical to the kagome lattice, but with half the length. Thereby, the twist operator for triangular lattice has the form
\begin{eqnarray}
\hat{O}=\exp \big[i\frac{2\pi}{N_1}\sum_{\vec{R}}(n_1+\frac{n_2}{2}) \hat{S}^z_{\vec{R}}\big]~,
\label{twistoperatortriangle}
\end{eqnarray}
with a notation identical to that used for the kagome lattice. Similarly, 
the OYA-like criterion for the triangular lattice is found to be
\begin{eqnarray}
\frac{Q_m}{2}\big(\frac{m}{m_s}-1\big)=n~.
\end{eqnarray}
This criterion offer a $1/3$-plateau as the simplest possibility via the enlargement of the magnetic unit cell, i.e., with $Q_m=3$ and $n=-1$, and is analogous to the FQH state with $\nu=1/3$.
This is consistent with predictions from numerical and experimental works
~\cite{chubokov1991quantum,PhysRevB.67.104431,PhysRevB.67.104431,PhysRevLett.102.137201,
takano2011self,yamamoto2016magnetization}. 
\section{Conclusions and Outlook}\label{consec}
\noindent
In conclusion, 
we have derived the twist operator 
for the kagome and triangular lattices.  Although the form of the twist operator is different from that for non-frustrated lattices, the non-commutativity between twist and translation operator is similar in the sense that it depends only on boundary unit cells. We have shown that the contribution from boundary spins 
leads to several possibilities for magnetization plateaux in frustrated systems. The plateaux are observed to be analogous to the integer and fractional quantum Hall states, 
%
offering insight into quantum liquid ground states with fixed numbers of singlets and spinons in the unit cell. While we have focussed on the case of $N_{2}$ being an odd integer in this work, some results can also be obtained for the case of $N_{2}$ being an even integer. For instance, for $Q_{m}=6$, we obtain magnetisation plateaux at $m/m_{s}=0, 1/3$ and $2/3$.
\par\noindent 
There are several interesting directions that are opened by our work. The first involves an investigation of whether the ground state wavefunctions we have obtained for some of the non-trivial magnetisation plateaux correspond to novel topological field theories. For instance, we have recently shown from a renormalisation group analysis that an effective Hamiltonian can be obtained for a quantum spin liquid phase of the Heisenberg quantum antifferomagnet corresponding to the $m/m_{s}=1/3$ plateau in the kagome lattice~\cite{pal2019}. This effective Hamiltonian was reached by the condensation of $SU(2)$ symmetric quantum fluctuations, suggesting that the problem can likely be studied in terms of a a $SU(2)$ non-Abelian lattice gauge theory on the kagome lattice  associated  with such quantum fluctuations~\cite{RevModPhys.51.659}. A continuum version of such a gauge theory is obtained from a fermionic non-linear sigma model of massive Dirac fermions in $(2+1)$ dimensions coupled to a SU(2) order parameter~\cite{abanov2000theta}, and found to lead to a quantum disordered ground state protected by a dynamically generated mass gap. Further, the theory is topolgical in nature, possessing a topological Hopf term in the effective action. It appears relevant, therefore, to investigate whether any the ground state wavefunction obtained by us for the plateau at $m/m_{s}=1/3$ in this work could be that for the quantum spin liquid ground state of Ref.\cite{pal2019}.
\par\noindent
In a recent work~ \cite{PhysRevB.100.104421}, the formalism developed here has been extended to the search for magnetization plateaus in other frustrated lattices, e.g., the pyrochlore in 3D. Any results obtained from a twist-operator based approach can likely provide considerable assistance in the experimental search for quantum spin liquids currently being sought in magnetic materials with frustrated geometries. Finally, we hope that this work will also motivate the search for plateaus that correspond to fractional values of the parameter $n$, in analogy with the fractional quantum Hall effect.
\section*{Acknowledgements} 
The authors thank S. Pujari, S. Patra, A. Panigrahi, R. K. Singh and G. Dev Mukherjee for several enlightening discussions. 
S. Pal and A. Mukherjee acknowledge CSIR, Govt. of India and IISER Kolkata for financial support. S. L. thanks the DST, Govt. of India for funding through a Ramanujan Fellowship (2010-2015) during which this project was initiated.
\appendix
\section{Energy cost of the Twisted state}\label{EnergyCost}
Here we present the calculation for energy different between the ground state ($|\psi_0\rangle$) and the twisted state ($|\psi_1\rangle$) generated due the application of twist operator on the ground state i.e. $|\psi_1\rangle=\hat{O}|\psi_0\rangle $
\begin{eqnarray}
\langle\psi_1|H|\psi_1 \rangle=\langle\psi_0|\hat{O}^{-1} H \hat{O}|\psi_0 \rangle~,
\label{DeltaE}
\end{eqnarray}
where the twist operator is defined by
\begin{eqnarray}
\hat{O}&=&\exp \big[i\frac{2\pi}{N_1}\Big(\sum_{\vec{r}}(n_1+\frac{n_2}{2}) \hat{S}^z_{\vec{r}}\nonumber\\
&&+\sum_{\vec{R}}(\frac{1}{4} \hat{S}^z_{\vec{R},b}+\frac{1}{2} \hat{S}^z_{\vec{R},c})\Big)\big]~.
\label{twistoperator1}
\end{eqnarray}
The meaning of various symbols is as defined in the main text. Using the following operator identities~\cite{lieb1961two} 
\begin{eqnarray}
\hat{O}^{-1} S^x_{\vec{R},a}\hat{O} &=& S^x_{\vec{R},a} \cos A+S^y_{\vec{R},a} \sin A~, \nonumber \\
\hat{O}^{-1} S^y_{\vec{R},a}\hat{O} &=& -S^x_{\vec{R},a} \sin A+S^y_{\vec{R},a} \cos A~, \nonumber \\
\hat{O}^{-1} S^z_{\vec{R},a}\hat{O}&=& S^z_{\vec{R},a}~,
\end{eqnarray}
\begin{eqnarray}
\hat{O}^{-1} S^x_{\vec{R},b}\hat{O} &=& S^x_{\vec{R},b} \cos B+S^y_{\vec{R},b} \sin B~, \nonumber \\
\hat{O}^{-1} S^y_{\vec{R},b}\hat{O} &=& -S^x_{\vec{R},b} \sin B+S^y_{\vec{R},b} \cos B~, \nonumber \\
\hat{O}^{-1} S^z_{\vec{R},b}\hat{O}&=& S^z_{\vec{R},b}~,
\end{eqnarray}
\begin{eqnarray}
\hat{O}^{-1} S^x_{\vec{R},c}\hat{O} &=& S^x_{\vec{R},c} \cos C+S^y_{\vec{R},c} \sin C~, \nonumber \\
\hat{O}^{-1} S^y_{\vec{R},c}\hat{O} &=& -S^x_{\vec{R},c} \sin C+S^y_{\vec{R},c} \cos C~, \nonumber \\
\hat{O}^{-1} S^z_{\vec{R},c}\hat{O}&=& S^z_{\vec{R},c}~,
\end{eqnarray}
where $a, b, c$ are the three sublattices of the Kagome lattice, we find the angles 
\begin{eqnarray}
A =\frac{2\pi}{N_1}(n_1+\frac{n_2}{2})~,&& ~B=\frac{2\pi}{N_1}(n_1+\frac{n_2}{2}+\frac{1}{4})\nonumber\\
\text{and}~~ C =\frac{2\pi}{N_1}&&(n_1+\frac{n_2}{2}+\frac{1}{2})~.
\end{eqnarray}
\begin{widetext}
Thus, we have
\begin{eqnarray}
\langle\psi_1|H|\psi_1 \rangle &=& \langle\psi_0|H|\psi_0\rangle+\langle\psi_0|[(\cos \frac{2\pi}{4N_1}-1)J\sum_{\vec{R}}(S^x_{\vec{R},a}S^x_{\vec{R},b}+S^y_{\vec{R},a}S^y_{\vec{R},b})\nonumber\\
&& +(\cos \frac{2\pi}{2N_1}-1)J\sum_{\vec{R}}(S^x_{\vec{R},a}S^x_{\vec{R},c}+S^y_{\vec{R},a}S^y_{\vec{R},c})\nonumber\\
&& +(\cos \frac{2\pi}{4N_1}-1)J\sum_{\vec{R}}(S^x_{\vec{R},b}S^x_{\vec{R},c}+S^y_{\vec{R},b}S^y_{\vec{R},c})]|\psi_0\rangle\\
&=& \langle\psi_0|H|\psi_0\rangle +\frac{N_{1}N_{2}}{2}\alpha J\left[ 2(1-\cos(\frac{2\pi}{4N_{1}})) + (1-\cos(\frac{2\pi}{2N_{1}}))\right]\nonumber\\
&\simeq & \langle\psi_0|H|\psi_0\rangle + \frac{N_{2}}{N_{1}}\frac{3\pi^{2}}{8}\alpha J + \mathcal{O}(N_1^{-3})~,
\end{eqnarray}
\end{widetext}
where $J$ denotes the spin exchange constant and the lattice constant (denoted by $\delta$ in the main manuscript) has been set to unity. In the fourth line, we have defined $N_{1}N_{2}\frac{\alpha}{2}=\langle\psi_{0}|\sum_{\vec{R}}(S^x_{\vec{R},i}S^x_{\vec{R},j}+S^y_{\vec{R},i}S^y_{\vec{R},j}) |\psi_{0}\rangle$~,~$(i,j)\in(a,b,c)$, $i\neq j$, as the ground state is a singlet of total spin, possessing rotational as well as translational invariances; it is thus expected to have a spin stiffness of equal expectation value in all spatial directions.

Further, we have expanded the cosine functions in the last line to leading order in $(1/N_{1})$. The factor $0\leq \alpha\leq 1$ denotes the renormalisation of the spin stiffness ($\rho=3\pi^{2}\alpha J/8N_{1}^{2}$), and is expected to vanish ($\alpha\to 0$) in a symmetry-preserved spin liquid~\cite{misguich2002degeneracy,PhysRevB.69.104431}. 
In this regard, we have also demonstrated recently from a RG analysis~\citep{pal2019} that the twist operator presented here is responsible for the formation of the spectral gap that protects the $1/3$ magnetization plateau ground state of the $S=1/2$ HKA on the kagome lattice.
For instance, in a gapped spin liquid displaying topological order, one finds~\cite{PhysRevB.69.104431,PhysRevLett.81.4484}
\begin{equation}
\alpha (L_{\hat{a}_{1}})\sim e^{-L_{\hat{a}_{1}}/\xi}~,~
\end{equation}
where $L_{\hat{a}_{1}}=2\delta N_{1}$ is the length along the twist direction ($\hat{a}_{1}$), $\delta$ is the lattice constant and $\xi$ denotes the correlation length. Thus, for isotropic ($N_{2}/N_{1})\sim \mathcal{O}(1)$) spin liquid states in two spatial dimensions, the vanishing of the spin stiffness $\rho$ (due to the vanishing of $\alpha$) leads to $\langle\psi_1|H|\psi_1 \rangle\to \langle\psi_0|H|\psi_0\rangle$. This ensures that the LSM theorem (based on the twist operator $\hat{O}$) is applicable for the study of spin liquid ground states in Heisenberg quantum antifferomagnets defined on geometrically frustrated lattices in two spatial dimensions. 
It is important to note that for zero-external magnetic field as the ground state $|\psi_0\rangle$ is a singlet of total spin, and therefore rotationally invariant, the expectation value of current-like terms (i.e., $S^x_{\vec{R},a} S^y_{\vec{R},b} - S^y_{\vec{R},a} S^x_{\vec{R},b}$ etc.) vanishes~\cite{auerbach2012interacting}. 
Such terms are also expected to have vanishing expectation values for the $U(1)$-symmetric plateau ground states at finite external field, as they are eigenstates of the total $S^{z}$ protected by a gap. Indeed, it can be shown from effective field theory and renormalisation group (RG) methods~\cite{tanaka2009geometric} that, in the presence of magnetic field, the gap responsible for the plateau is robust against such current-like terms.

\section{Details of the calculation for the LSM-like theorem for kagome lattice}\label{LSMcalculation}
For PBC along $\hat{a}_1$ direction, we have
\begin{widetext}
\begin{eqnarray}
\hat{T}_{\hat{a}_1}\hat{O}_a\hat{T}^\dagger_{\hat{a}_1}&=&\exp[i\frac{2\pi}{N_1}\sum_{n_2}\{ (1+\frac{n_2}{2})\hat{S}^z_{2,n_2,a}+ (2+\frac{n_2}{2})\hat{S}^z_{3,n_2,a}+...+(N_1+\frac{n_2}{2})\hat{S}^z_{N_1+1,n_2,a} \}] \nonumber\\
&= &\hat{O}_a \exp[-i\frac{2\pi}{N_1}(\hat{S}^z_{\text{Tot}})_a]\exp[i2\pi\sum_{n_2}\hat{S}^z_{1,n_2,a}]~.
\end{eqnarray} 
Similarly, we find 
\begin{eqnarray}
\hat{T}_{\hat{a}_1}\hat{O}_b\hat{T}^\dagger_{\hat{a}_1} = \hat{O}_b \exp[-i\frac{2\pi}{N_1}(\hat{S}^z_{\text{Tot}})_b]\exp[i2\pi\sum_{n_2}\hat{S}^z_{1,n_2,b}]~,
\end{eqnarray}
and 
\begin{eqnarray}
\hat{T}_{\hat{a}_1}\hat{O}_c\hat{T}^\dagger_{\hat{a}_1} = \hat{O}_c \exp[-i\frac{2\pi}{N_1}(\hat{S}^z_{\text{Tot}})_c]\exp[i2\pi\sum_{n_2}\hat{S}^z_{1,n_2,c}]~.
\end{eqnarray}
Then, bringing all these relations together, we find
\begin{eqnarray}
\hat{T}_{\hat{a}_1}\hat{O}\hat{T}^\dagger_{\hat{a}_1} &=& \hat{T}_{\hat{a}_1}\hat{O}_a\hat{T}^\dagger_{\hat{a}_1} \hat{T}_{\hat{a}_1}\hat{O}_b\hat{T}^\dagger_{\hat{a}_1} \hat{T}_{\hat{a}_1}\hat{O}_c\hat{T}^\dagger_{\hat{a}_1}~~~~ (\because \hat{T}^\dagger_{\hat{a}_1}\hat{T}_{\hat{a}_1}=\mathbb{I})\nonumber \\
&=& \hat{O}_a\hat{O}_b\hat{O}_c \exp[-i\frac{2\pi}{N_1}\{ (\hat{S}^z_{\text{Tot}})_a+(\hat{S}^z_{\text{Tot}})_b+(\hat{S}^z_{\text{Tot}})_c\}]\exp[i2\pi\sum_{n_2}( \hat{S}^z_{1,n_2,a}+\hat{S}^z_{1,n_2,b}+\hat{S}^z_{1,n_2,c})]\nonumber \\
&=& \hat{O}\exp\big[-i\frac{2\pi}{N_1}(\hat{S}^z_{Tot}-N_1 N_2 \hat{S}^z_{\triangle})\big]~,
\label{LSMcriterian1}
\end{eqnarray}
\end{widetext}
where the total magnetization is given by $\hat{S}^z_{Tot}=\sum_{\vec{r}}\hat{S}^z_{\vec{r}}$, and $N_{2}\hat{S}^z_\triangle$ 
is the $z$-component of the vector sum of all spins within the $N_{2}$ unit cells lying on a line along $\hat{a}_{2}$. 

\bibliography{kagometwistbib}
\end{document}